\pdfoutput=1
\documentclass[11pt,twoside,a4paper,cmspaper,final,collab]{cms-tdr}
\def\svnVersion{9f2ec50}\def\svnDate{2025/06/18}\def\cmsCernNoTag{CERN-EP-2024-345}\def\cmsCernDate{\today}\def\cmsMessage{Published in the Journal of High Energy Physics as \href{http://dx.doi.org/10.1007/JHEP06(2025)120}{\doi{10.1007/JHEP06(2025)120}.}}
\begin{document}\cmsNoteHeader{HIN-24-010}

\newcommand{\PbPb}{\ensuremath{\text{PbPb}}\xspace}
\newcommand{\pp}{\ensuremath{\Pp\Pp}\xspace}
\newcommand{\roots}{\ensuremath{\sqrt{s}}\xspace}
\newcommand{\DRaxis}{\ensuremath{\Delta\text{j}}\xspace}
\newcommand{\JEWEL}{\textsc{jewel}\xspace}
\newlength\cmsTabSkip\setlength{\cmsTabSkip}{1ex}

\cmsNoteHeader{HIN-24-010}

\title{Search for medium effects using jet axis decorrelation in inclusive jets from PbPb collisions at \texorpdfstring{$\sqrtsNN = 5.02\TeV$}{sqrtsNN = 5.02 TeV}}

\author*[cern]{The CMS Collaboration}

\date{\today}

\abstract{The jet axis decorrelation in inclusive jets is studied using lead-lead (PbPb) collisions at a center-of-mass energy per nucleon pair of 5.02 TeV. The jet axis decorrelation is defined as the angular difference between two definitions of the jet axis. It is obtained by applying two recombination schemes on all the constituents of a given jet reconstructed by the anti-\kt sequential algorithm with a distance parameter of $R = 0.4$. The data set, corresponding to an integrated luminosity of 0.66\nbinv, was collected in 2018 with the CMS detector at the CERN LHC. The jet axis decorrelations are examined across collision centrality selections and intervals of jet transverse momentum. A centrality dependent evolution of the measured distributions is observed, with a progressive narrowing seen in more central events. This narrowing could result from medium-induced modification of the internal jet structure or reflect color charge effects in energy loss. This new measurement probes jet substructure in previously unexplored kinematic domains and show great promise for providing new insights on the color charge dependence of energy loss to jet-quenching models.}

\hypersetup{
pdfauthor={CMS Collaboration},
pdftitle={Search for medium effects using jet axis decorrelation in inclusive jets from PbPb collisions at 5.02 TeV},%
pdfsubject={CMS},
pdfkeywords={CMS, heavy ion, jet axis}}

\maketitle 

\section{Introduction}
\label{sec:intro}
Jets have been used as versatile tools for probing the properties of quark-gluon plasma (QGP) created in high-energy collisions~\cite{Apolinario:2022vzg, Cunqueiro:2021wls}. Jets are collimated streams of particles that form when energetic quarks and gluons (partons) produced in high-momentum-transfer (hard) processes fragment into lower-energy (soft) partons and subsequently hadronize into particles. During the evolution of the jet shower, the outgoing partons can interact with the QGP, resulting in a range of physical phenomena collectively known as ``jet quenching.'' Experimental evidence of jet quenching is provided by many observables at the BNL RHIC~\cite{PHENIX:2001hpc, STAR:2002ggv} and the CERN LHC~\cite{CMS:2021vui, CMS:2021nhn, ATLAS:2023hso, ALICE:2023waz}. Jet-medium interactions in the QGP modify the jet shower properties as compared to jets produced in collisions of elementary particles. Various jet substructure observables are employed to study radiation patterns inside jets and the effects of QGP on those distributions. There have been numerous experimental measurements indicating a modification of jet substructure in heavy ion collisions as compared to proton-proton (\pp) collisions~\cite{CMS:2024girth, CMS:HIN-21-019, ALICE:angular}. A recent review of CMS measurements on the subject can be found in Ref.~\cite{CMS:2024krd}.

Energy scale differences between the initial parton and final-state shower constituents, as well as nonperturbative hadronization effects, pose challenges for theoretical predictions of jet quenching. Additional complications arise from the response of the medium to the propagating jet and how this response could further modify the jet shower. Perturbative and nonperturbative contributions, which arise from different energy scales and processes, are treated differently in various theoretical approaches, with a greater degree of variation and uncertainty for the nonperturbative components. It is therefore critical to map out jet modification at different scales and to probe the internal jet substructure with observables that have different sensitivities to the perturbative and nonperturbative domains.

In this paper, a search for possible modifications to jet substructure is conducted using measurements of the jet axis decorrelation between two different jet axes within a given jet, which are expected to have different sensitivities to soft-gluon radiation~\cite{angelJetaxis}. The jet axis decorrelation, \DRaxis, is defined as the angular separation between the axes found using the traditional anti-\kt energy-weighted (E-scheme) recombination and the winner-take-all (WTA) recombination schemes~\cite{angelJetaxis}:
\begin{equation}
\DRaxis = \sqrt{{(\eta^{\text{E-scheme}} - \eta^{\text{WTA}})^{2} + (\phi^{\text{E-scheme}} - \phi^{\text{WTA}})^{2}}},
\label{eq:DRaxis}
\end{equation}
where each jet axis direction is characterized by its pseudorapidity ($\eta$) and azimuthal angle ($\phi$). In the E-scheme recombination algorithm, the four-vector sum of constituents determines the jet axis at each clustering step. This definition includes contributions from soft constituents and provides the momentum-averaged direction of the energy flow. On the other hand, the WTA recombination scheme follows the direction of the hardest prong at each clustering iteration and is more sensitive to collinear radiation inside the jet~\cite{angelJetaxis}. The WTA axis is insensitive to soft radiation at leading order, allowing for a more direct interpretation in terms of perturbative quantum chromodynamics (QCD) calculations. Since these two algorithms have different sensitivities to soft radiation, studies of axis decorrelations can provide information about radiative energy loss in QGP. These decorrelations are also sensitive to the details of the parton shower and can reflect modifications of fragmentation patterns as well as their dependence on the parton flavor.

Experimental measurements of \DRaxis provide significant advantages for tuning the parton shower in event generators and for gaining insight on medium effects in heavy ion collisions because of the availability of analytical calculations at next-to-leading logarithmic accuracy~\cite{angelJetaxis}. Successful applications for the former can be found in comparisons of numerical calculations for \pp collisions at a center-of-mass energy $\roots = 13\TeV$ with \PYTHIA 8.2 simulations, where excellent agreement has been demonstrated for transverse momenta (\pt) above 100\GeV~\cite{angelJetaxis}. However, at lower \pt, discrepancies have been observed, likely attributable to nonperturbative effects and power-suppressed contributions, which arise from soft radiation and hadronization processes that become significant at low energy scales~\cite{angelJetaxis}.

Recently, the ALICE Collaboration measured jet-axis decorrelations in lead-lead (\PbPb) and \pp collisions at a center-of-mass energy per nucleon pair (\sqrtsNN) of 5.02\TeV using E-scheme and WTA axes with jets reconstructed from charged particle tracks~\cite{ALICE:angular, ALICE:angularpp}. The measured distributions for jets in the range $40<\pt<80\GeV$ showed a narrowing of the \DRaxis distributions in \PbPb collisions compared to the \pp reference. These measurements were primarily conducted in a kinematic region where nonperturbative effects, such as contributions from the medium response and underlying event fluctuations, are expected to be more significant. Similarly, the CMS Collaboration also measured jet axis decorrelations in \PbPb and \pp collisions at $\sqrtsNN = 5.02\TeV$ using isolated photon-tagged jet events in the $30<\pt<100\GeV$ range~\cite{CMS:HIN-21-019}. There, a similar narrowing of the \DRaxis distribution in \PbPb compared to \pp collisions was seen in the momentum range of $60<\pt<100\GeV$.

This paper presents a study of the jet axis decorrelation of anti-\kt jets with a distance parameter of $R = 0.4$ in \PbPb collisions at $\sqrtsNN = 5.02\TeV$, using data recorded by the CMS experiment in 2018. Measurements are carried out in four \pt intervals ranging from 120 to 300\GeV and in four centrality intervals from 0 to 80\%, where centrality is defined as the fraction of the total nucleus-nucleus cross section, with 0\% corresponding to the maximum overlap of the colliding nuclei. Measurements are corrected for detector resolution effects by employing an unfolding procedure. The sensitivity of the \DRaxis distribution shape to the type of parton initiating the jet shower is also employed to explore the potential impact of color charge effects on energy loss, and several model scenarios are examined. Tabulated results are provided in the HEPData record for this analysis~\cite{hepdata}.

\section{Experimental setup and data sample}
\label{sec:exp_data}
The central feature of the CMS apparatus is a superconducting solenoid of 6\unit{m} internal diameter, providing a magnetic field of 3.8\unit{T}. Within the solenoid volume are a silicon pixel and strip tracker, a lead tungstate crystal electromagnetic calorimeter, and a brass and scintillator hadron calorimeter, each composed of a barrel and two endcap sections. Two hadron forward (HF) steel and quartz-fiber calorimeters complement the barrel and endcap detectors, providing coverage to $\abs{\eta}<5.2$. The HF calorimeters are segmented into multiple $0.175{\times}0.175$ ($\Delta\eta{\times}\Delta\phi$) ``towers'', where $\phi$ is the azimuthal angle in radians. Events of interest are selected using a two-tiered trigger system. The first level, composed of custom hardware processors, uses information from the calorimeters and muon detectors to select events at a rate of around 100\unit{kHz}~\cite{Sirunyan:2020zal}. The second level, known as the high-level trigger, consists of a farm of processors running a version of the full event reconstruction software optimized for fast processing and reducing the event rate to around 1\unit{kHz} before data storage~\cite{CMS:trigger2017}. A detailed description of the CMS detector, together with a definition of the coordinate system used and the relevant kinematic variables, can be found in Refs.~\cite{CMSexperiment, CMS:2023gfb}.

This analysis uses approximately 7 million events, corresponding to an integrated luminosity of 0.66\nbinv~\cite{CMS:LUM-17-003, CMS:LUM-17-004, CMS:LUM-18-002}, that were recorded in 2018 using high-level triggers requiring at least one jet with $\pt > 80\GeV$. Background events due to beam-gas interactions and nonhadronic collisions are removed offline by applying the criteria described in Ref.~\cite{Khachatryan:2016odn}. Events are required to have at least one primary interaction vertex reconstructed using two or more tracks~\cite{Collaboration_2014} within a distance of 15\unit{cm} from the center of the nominal interaction point along the beam axis. Each event must have at least two calorimeter towers in each HF detector, with energy deposits exceeding 4\GeV per tower, effectively removing the ultraperipheral collision events. The shapes of the clusters in the pixel detector are required to be compatible with those expected in \PbPb collisions to prevent contamination from events with multiple collisions~\cite{JHEP022010}. The collision centrality is determined using the total sum of transverse energy (\et) from calorimeter towers in the HF region (covering $2.9<\abs{\eta}<5.2$). The measured HF energy distribution is used to divide the event sample into bins, each representing 0.5\% of the total nucleus-nucleus hadronic interaction cross section. A detailed description of centrality determination can be found in Ref.~\cite{Chatrchyan:2011sx}.

A sample of simulated events is used to correct reconstruction and detector resolution effects and to construct template distributions for jets from different types of partons in modeling studies. The \PYTHIA 8.226~\cite{pythia82} event generator with tune CP5~\cite{pythiaCP5tune} and NNPDF3.1 parton distribution functions at next-to-next-to-leading order~\cite{nnpdf31} is used to describe the hard scattering, parton showering, and hadronization processes for the interaction producing jets of interest. These events are embedded into simulated minimum bias \PbPb collisions produced with the \HYDJET 1.9~\cite{Lokhtin:2005px} event generator to account for the soft underlying event produced in \PbPb collisions. The \HYDJET generator is tuned to mirror global \PbPb event properties such as the charged-hadron \pt spectrum and particle multiplicity. The combined sample of hard \PYTHIA interactions and soft \HYDJET underlying event is referred to as the {\PYTHIA}+\HYDJET sample. Simulation of the CMS detector response is performed using the \GEANTfour~\cite{GEANT4} toolkit.

In \PbPb collisions, the frequency of jet production is higher in central events due to the increased number of binary nucleon-nucleon collisions per nuclear interaction. Conversely, the {\PYTHIA}+\HYDJET sample exhibits a uniform distribution of jets (from {\PYTHIA}) across centrality bins. Therefore, a centrality-based weighting is applied to the minimum bias MC sample to align its centrality distribution with the jet-triggered \PbPb data. Another weighting step is performed to match the distributions of the primary vertex position along the beam direction in the simulated and data samples.

\section{Jet reconstruction}
\label{sec:rec_jet}
In \PbPb collisions, jets are reconstructed using the anti-\kt algorithm~\cite{Cacciari:2008gp} with a distance parameter of $R = 0.4$, implemented in the \FASTJET framework~\cite{Cacciari:2011ma}. A particle-flow (PF) algorithm, optimized to combine information from various elements of the CMS detector, is used to reconstruct leptons, photons, and charged and neutral hadrons~\cite{CMS:2017yfk}. These PF candidates are the input for reconstructing the jets analyzed in this study.  The constituent subtraction method~\cite{constituentSubtraction} is employed to subtract the contributions of the underlying event.  Charged-hadron PF candidates are assumed to have the pion mass. The jets are clustered using the E-scheme recombination scheme~\cite{Cacciari:2011ma}, wherein particle pairs are iteratively combined to form pseudo-jets. These pseudo-jets are objects that combine particles or other pseudo-jets. The direction of the new pseudo-jet is determined by the vector sum of the four-momenta of its constituent particles. When using E-scheme recombination scheme, the direction of the resulting four-vector is the E-scheme jet axis, expressed in $(\eta,\phi)$ coordinates.  To calculate the other axis relevant to this work, the WTA axis, the jet is reclustered using the winner-take-all recombination scheme~\cite{wta1,wta2} combined with the anti-\kt algorithm and the same input PF constituents. During each iteration, the direction of the new pseudo-jet aligns with that of the highest-\pt particle or pseudo-jet being combined. Consequently, the WTA axis typically coincides with the direction of the hardest jet constituent.

Jet energy corrections are derived from simulation studies so that the average measured energy of jets becomes identical to that of particle-level jets. In situ measurements of the momentum balance using dijet events from \pp collisions are used to determine any residual differences between the jet energy scale (JES) in data and simulation, and appropriate corrections are made~\cite{CMS:2016lmd}. Additional selection criteria are applied to each jet to remove jets potentially dominated by instrumental effects or reconstruction failures~\cite{CMS:2020ebo}.

After jet reconstruction and energy correction, all selected jets are required to have $\pt > 120\GeV$ and $\abs{\eta} < 1.6$. In this kinematic range, the jet trigger is fully efficient, and the jet energy response and resolution are optimal. The study is performed in four intervals of jet \pt, specifically $120 < \pt < 150\GeV$, $150 < \pt < 190\GeV$, $190 < \pt < 230\GeV$, and $230 < \pt < 300\GeV$.

\section{Analysis method}
\label{sec:ana}
The jet axis decorrelation between the E-scheme and WTA axes is calculated according to Eq.(~\ref{eq:DRaxis}). The measured \DRaxis distributions are normalized by the total number of jets within each jet \pt interval, denoted as $\mathrm{N}(\pt^{\text{jet}})$, yielding the following expression:
\begin{equation*}
\frac{1}{\mathrm{N}(\pt^{\text{jet}})}\frac{\rd\mathrm{N}(\pt^{\text{jet}})}{\rd\DRaxis}.
\end{equation*}
\noindent Measurements are unfolded to the particle level to correct detector resolution effects and underlying event fluctuations, facilitating comparisons with other experiments and theoretical predictions. The unfolding process employs the iterative D'Agostini's unfolding method~\cite{DAgostini1995487, Richardson, Lucy} with early stopping, which is implemented in the \textsc{RooUnfold} software package~\cite{Adye:2011gm} and requires a response matrix. This matrix is derived from simulated samples and captures the correspondence between the \DRaxis and jet \pt at both the generator and detector levels. To capture bin-migration effects, response matrices were constructed using the detector-level \pt in the range 80--500\GeV. Within the kinematic range studied, jet reconstruction is fully efficient even in the most central \PbPb collisions. Thus, the unfolding procedure does not include additional scaling factors for missed jets. Additionally, erroneous jet contributions from random background fluctuations were found to be negligible for jets above 120\GeV, so no corrections for this effect are applied.

The \DRaxis distributions in the simulated samples exhibit substantial shape differences between quark- and gluon-initiated jets across all \pt intervals investigated in this analysis. These generator-level distributions are subsequently used for template fitting to estimate the potential limits of color charge effects in jet energy loss.

\section{Systematic uncertainties}
\label{sec:sysun}
The systematic uncertainties are determined by individually varying each analysis criterion, as discussed below. 

Event centrality is determined in the data by the \et sum in the HF calorimeters, which involves inherent uncertainties. The boundaries of \et sums in HF for each centrality interval are varied according to these uncertainties to assess their impact on the final result. The resulting difference is taken as a systematic uncertainty.

The systematic uncertainty associated with jet energy resolution (JER) arises from possible discrepancies in the JER between data and simulation. In the nominal analysis, these differences are accounted for by applying a jet-$\eta$ dependent smearing to the jet \pt in the simulation when constructing response matrices. To estimate the systematic uncertainty, new response matrices are generated by adjusting the smearing amount (both up and down) to reflect the uncertainty in the data-to-simulation discrepancy. The unfolding procedure is then performed using these modified response matrices. Additionally, a centrality-dependent systematic uncertainty due to JER is examined to account for potential differences in the background energy density between data and simulation. This uncertainty is estimated by shifting the centrality in the simulation up and down by 2\%, which accounts for the observed particle density variations between data and MC. The systematic uncertainty due to the JES correction is calculated by shifting the jet \pt up and down in the data according to the JES uncertainties~\cite{CMS:2016lmd} and comparing the shifted and nominal results.

In the nominal jet reconstruction method, the constituent subtraction method~\cite{constituentSubtraction} is used to subtract the contributions of the underlying event from the jet energy. An alternative background subtraction technique, accounting for the effects of hydrodynamical flow via event-by-event Fourier fits to the $\phi$ distribution of PF candidates~\cite{CMS:2025ydi}, is used for the systematic evaluation. 

The optimal number of unfolding iterations is determined by analyzing the convergence behavior and the bias-variance trade-off through cross-unfolding different MC samples (unfolding reweighted MC samples using response matrices from nominal MC and vice versa). The iterative procedure is stopped when, for all bins, the magnitude of the bias becomes smaller than the statistical uncertainty, ensuring a balance between accuracy and overfitting. Based on this criterion, the optimal number of iterations is three. Results obtained using two and four iterations are compared to assess the systematic uncertainty associated with the iteration count. The response matrices used in the unfolding procedure are derived from the {\PYTHIA}+\HYDJET sample, which has limited statistical precision. The statistical uncertainty inherent in these response matrices can influence the unfolding process. To account for the resulting uncertainty, fifty alternative response matrices are constructed by randomly varying each response matrix bin content according to the statistical uncertainty of the nominal response matrix and repeating the unfolding process. The resulting systematic uncertainty is determined by calculating the root-mean-square value from the fifty unfolding outputs and evaluating the differences from the nominal values. To study the systematic effect of the response matrix populated from a specific MC sample, a new response matrix is obtained by varying the quark- and gluon-like fractions in the simulation based on limits obtained from data and performing the unfolding. The {\PYTHIA}+\HYDJET sample, where the jet \pt is weighted to match the data, is used to construct the nominal response matrices. The unfolding procedure is repeated with a new response matrix constructed using the unweighted jet \pt distribution to evaluate the uncertainty related to the prior shape. 

The energy scale uncertainty of PF candidates was checked by scaling the four-momenta of the charged hadron and photon PF candidates by $\pm$1\% and of neutral hadron PF candidates by $\pm$3\%~\cite{CMS:2024girth}. The impact of these variations was found to be negligible.

When multiple variations are considered for a single uncertainty source, the maximum deviation from the nominal value is used. The total systematic uncertainty is obtained by summing the uncertainties from each source in quadrature. In this analysis, systematic uncertainties are categorized as either correlated or uncorrelated.

The systematic uncertainties due to the JER and JES are treated as correlated sources, while all other uncertainties are treated as uncorrelated. The systematic variations due to JER and the JES affect the jet \pt, which is anti-correlated with \DRaxis. The dominant systematic uncertainty in this analysis is from the JES. The absolute contributions from each source, as well as the total systematic uncertainties, are summarized in Table~\ref{tab:systematics_DR} for \DRaxis distributions.

\begin{table}[htbp!]
  \centering
  \topcaption{Summary of absolute systematic uncertainties, averaged over the \DRaxis distributions for different centrality intervals. The range of uncertainty values reflects the maximum variation across different \pt intervals.}
  \label{tab:systematics_DR}
  \begin{tabular}{lcccccc}
    \hline
    \multirow{2}{*}{Uncertainty source}  &\multicolumn{4}{c}{Centrality} \\
    \cline{2-5}
    & 50--80\% & 30--50\% & 10--30\% & 0--10\%\\                                          
    \hline
    Centrality                           & 0.001--0.011 & 0.001--0.004 & 0.000--0.002 & 0.000--0.002\\
    JER                                  & 0.053--0.088 & 0.068--0.123 & 0.054--0.180 & 0.189--0.290\\          
    JES                                  & 0.304--0.574 & 0.355--0.443 & 0.407--0.487 & 0.423--0.498\\ 
    Background subtraction               & 0.006--0.041 & 0.023--0.048 & 0.037--0.068 & 0.067--0.091\\ 
    Iteration choice                     & 0.007--0.053 & 0.009--0.041 & 0.020--0.057 & 0.017--0.086\\  
    Response matrix event count          & 0.000--0.001 & 0.000--0.001 & 0.000--0.001 & 0.000--0.001\\ 
    Scaled flavor fraction in sim.       & 0.000--0.002 & 0.001--0.002 & 0.003--0.006 & 0.005--0.012\\   
    Prior choice                         & 0.001--0.002 & 0.001--0.002 & 0.001--0.003 & 0.006--0.013\\    [\cmsTabSkip]
    Total correlated uncertainty         & 0.316--0.581 & 0.362--0.459 & 0.443--0.490 & 0.463--0.534\\
    Total uncorrelated uncertainty       & 0.015--0.066 & 0.026--0.054 & 0.065--0.072 & 0.088--0.110\\ 
  \end{tabular}
\end{table} 

\section{Results and discussion}
\label{sec:results}
\begin{figure}[htbp!]
\centering
\includegraphics[width=0.99\textwidth]{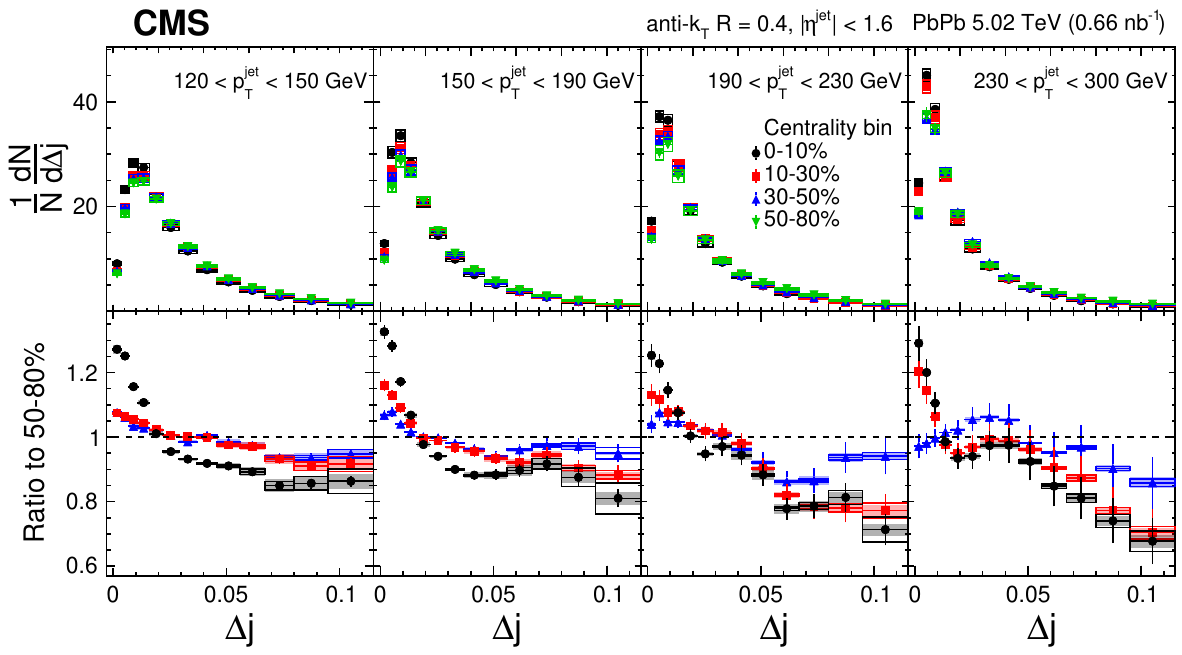}
\caption{The unfolded normalized \DRaxis distributions are measured across different centrality and \pt intervals. Each panel displays distributions for four centrality intervals within a specific \pt bin, with \pt ranges (indicated in each panel) increasing from left to right. In each panel, the black circles show measurements for 0--10\% centrality interval, the red squares for 10--30\%, and the blue-up and green-down triangles for 30--50\% and 50--80\% centrality intervals, respectively. The vertical solid lines represent the statistical uncertainties, while the rectangles and shaded areas represent the correlated and uncorrelated systematic uncertainties, respectively.}
\label{fig:unfoldedCorr}
\end{figure}

Figure~\ref{fig:unfoldedCorr} shows the unfolded normalized \DRaxis distributions in different jet \pt intervals ranging from $120 < \pt < 150\GeV$ to $230 < \pt < 300\GeV$.  Each plot consists of four distributions corresponding to four centrality intervals, distinguished by different marker colors. The lower panels show ratios of \DRaxis distributions in a given centrality range to that of the 50--80\% centrality interval. Correlated systematic uncertainties, the dominant sources of uncertainty in this analysis, are mitigated in the ratios by dividing each distribution by the corresponding distribution from the 50--80\% centrality interval. The \DRaxis distributions dip towards the smallest \DRaxis values ($\DRaxis\approx0$), indicating a low probability of finding both the E-scheme and WTA axes of a jet aligned in the same direction. These distributions peak between $0.01 < \DRaxis < 0.03$, and then decline sharply with increasing \DRaxis.

A progressive narrowing of the \DRaxis distributions, evident in the enhancement of small-\DRaxis yields and a compensating suppression at higher \DRaxis values, is observed from peripheral to central collisions across all \pt intervals. The narrowing is most significant in the most central events and could be attributed to modifications of the internal structure of the jet induced by interactions with the QGP medium. However, one could expect some broadening because of in-medium gluon radiation influencing the direction of the jet axes (particularly the E-scheme axis). It has also been argued in another substructure measurement that such narrowing in inclusive jets could be a consequence of a selection bias~\cite{CMS:2024girth}, with different amounts of energy loss experienced by jets with different internal structures due to, e.g., color coherence effects. Additionally, as higher \pt jets have narrower \DRaxis distributions compared to lower-\pt jets across all centrality intervals, the apparent narrowing could be an artifact of out-of-cone partonic energy loss affecting the JES. That is, the reconstruction of the quenched jet does not preserve the same correspondence to the initial parton energy as the jets that fragmented in the vacuum, causing a momentum selection bias that would be present in the heavy ion data.

While a direct comparison to the ALICE results~\cite{ALICE:angular} from the same collision system and energy is precluded by the differences in the technical aspects of jet reconstruction and jet axis definitions, the physics observations related to the narrowing of jet axis decorrelation are consistent between the two experiments within the overlapping kinematic range.

\begin{figure}[htbp!]
\centering
\includegraphics[width=0.99\textwidth]{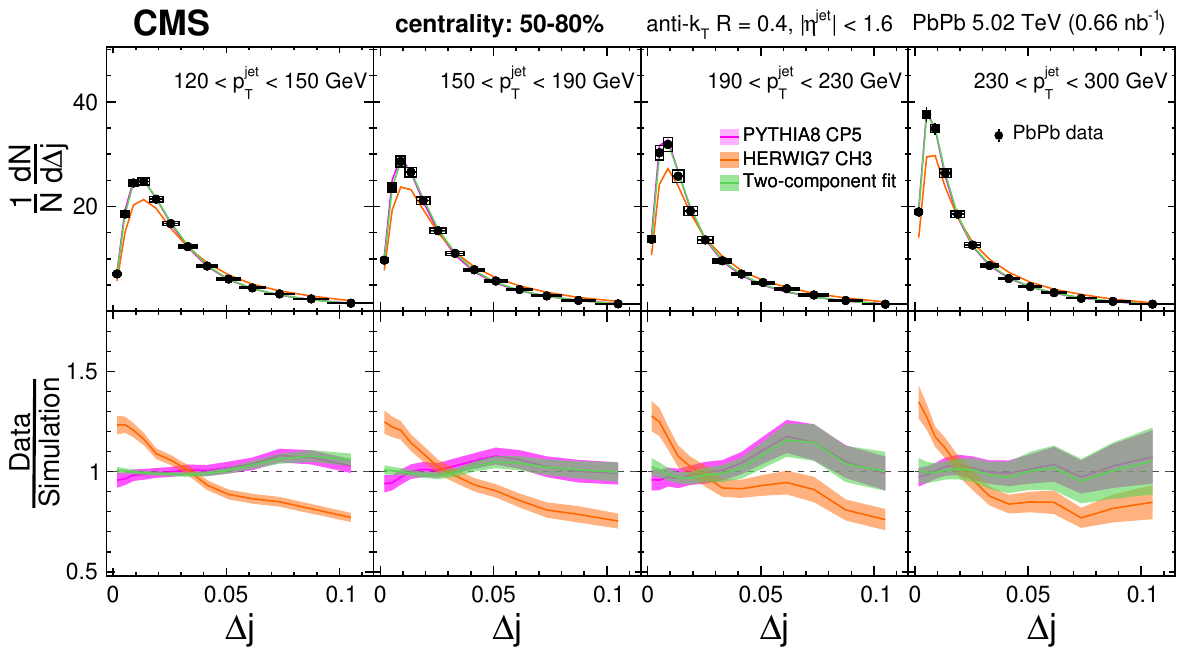}
\caption{The unfolded \DRaxis distributions (black circles) are compared with simulations from \PYTHIA, \HERWIG, and a two-component, represented as colored bands, for the 50--80\% centrality range across different \pt intervals. The vertical solid lines represent the statistical uncertainties, while the rectangles and shaded areas represent the correlated and uncorrelated systematic uncertainties, respectively. Only statistical uncertainties are shown for the simulated distributions.}
\label{fig:unfoldedCorr_ModelComp_Cent4}
\end{figure}

\begin{figure}[htbp!]
\centering
\includegraphics[width=0.99\textwidth]{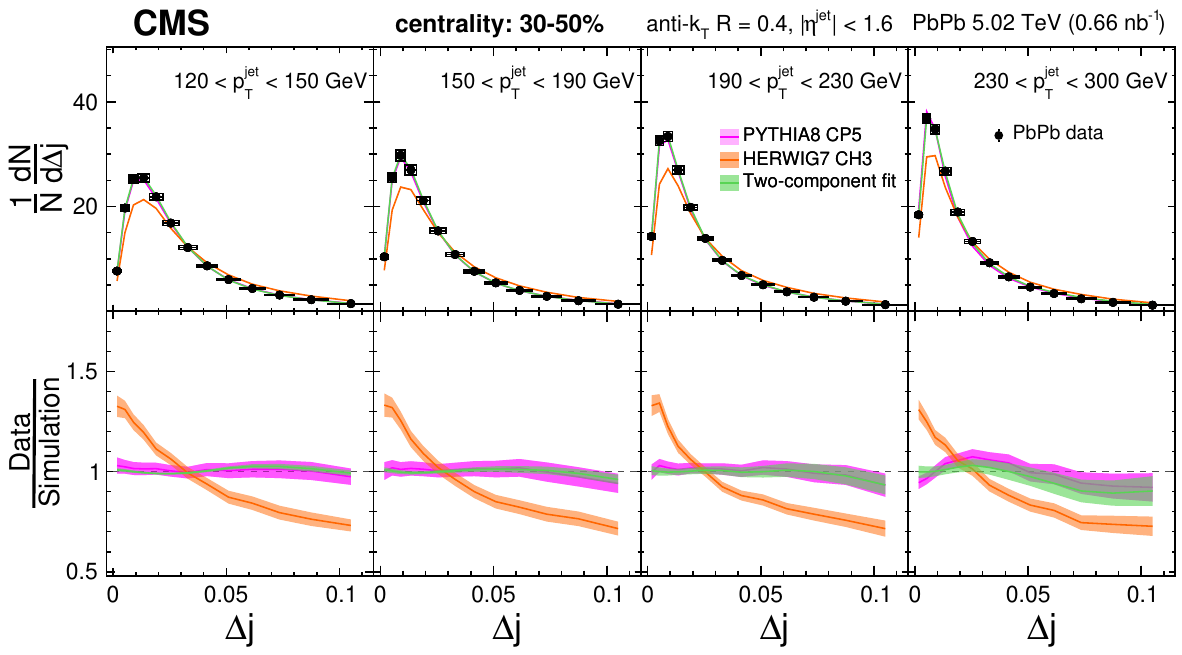}
\caption{The unfolded \DRaxis distributions (black circles) are compared with simulations from \PYTHIA, \HERWIG, and a two-component, represented as colored bands, for the 30--50\% centrality range across different \pt intervals. The vertical solid lines represent the statistical uncertainties, while the rectangles and shaded areas represent the correlated and uncorrelated systematic uncertainties, respectively. Only statistical uncertainties are shown for the simulated distributions.}
\label{fig:unfoldedCorr_ModelComp_Cent3}
\end{figure}

\begin{figure}[htbp!]
\centering
\includegraphics[width=0.99\textwidth]{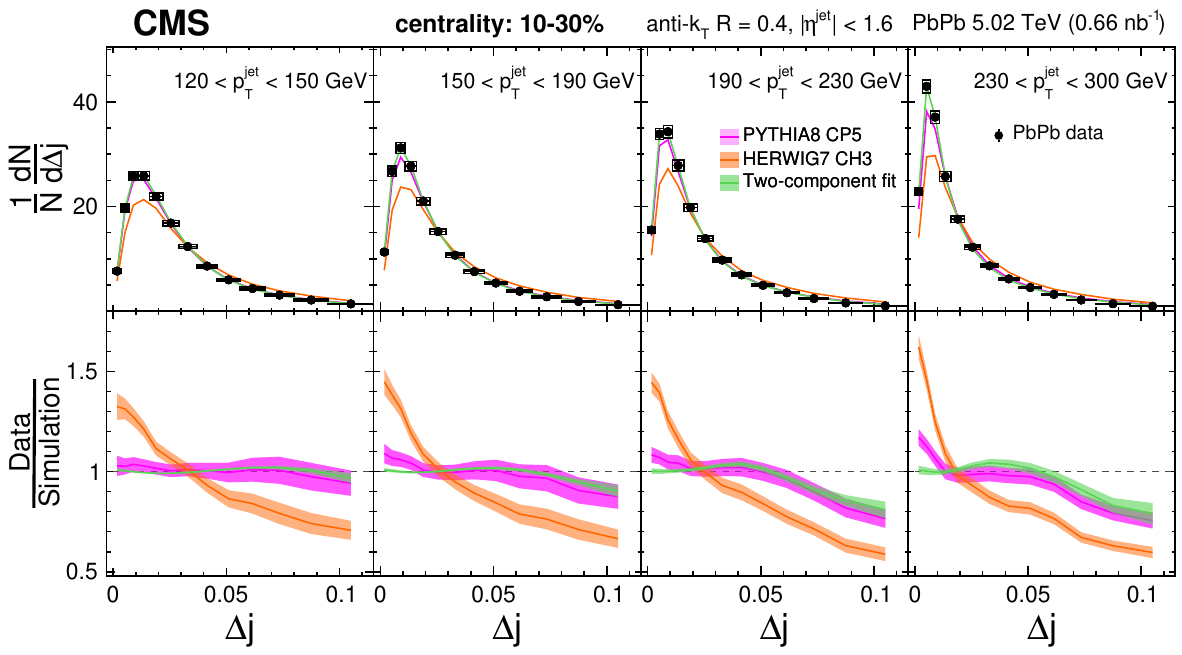}
\caption{The unfolded \DRaxis distributions (black circles) are compared with simulations from \PYTHIA, \HERWIG, and a two-component, represented as colored bands, for the 10--30\% centrality range across different \pt intervals. The vertical solid lines represent the statistical uncertainties, while the rectangles and shaded areas represent the correlated and uncorrelated systematic uncertainties, respectively. Only statistical uncertainties are shown for the simulated distributions.}
\label{fig:unfoldedCorr_ModelComp_Cent2}
\end{figure}

\begin{figure}[htbp!]
\centering
\includegraphics[width=0.99\textwidth]{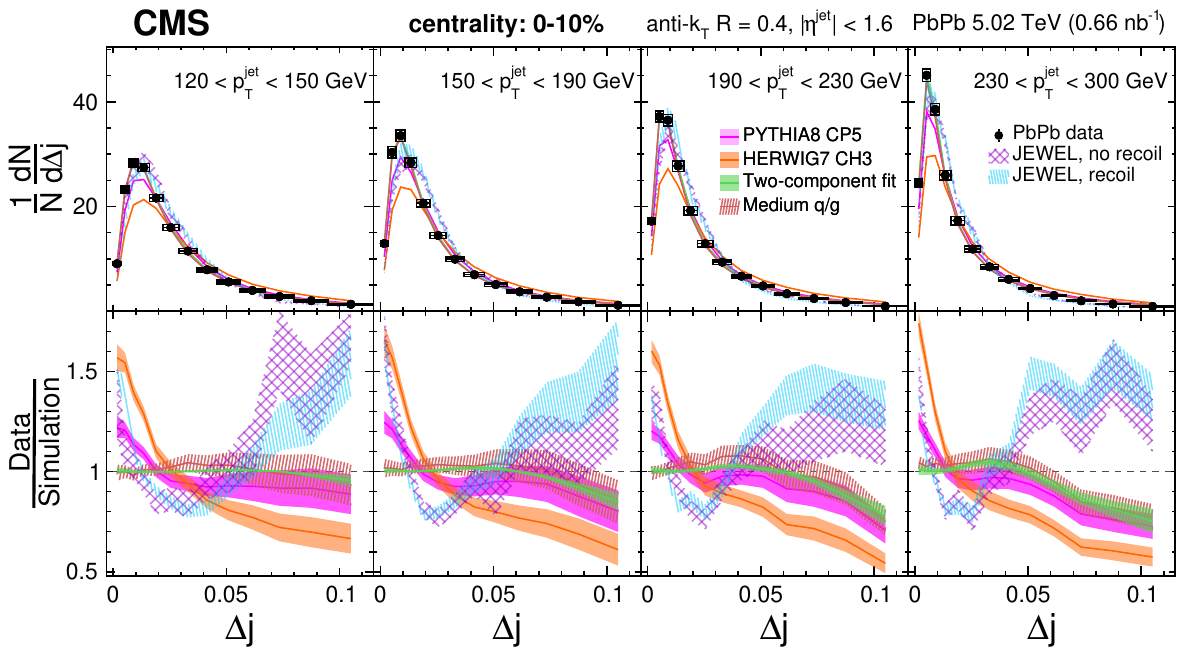}
\caption{The unfolded \DRaxis distributions (black circles) are compared with simulations from \PYTHIA, \HERWIG, and a two-component (represented as colored bands), as well as medium q/g and \JEWEL predictions (represented as hatched lines) for the 0--10\% centrality range across different \pt intervals. The vertical solid lines represent the statistical uncertainties, while the rectangles and shaded areas represent the correlated and uncorrelated systematic uncertainties, respectively. Only statistical uncertainties are shown for the simulated distributions.}
\label{fig:unfoldedCorr_ModelComp_Cent1}
\end{figure}

The new results are compared with the {\PYTHIA}8 MC generator using the CP5 tune~\cite{pythia82, pythiaCP5tune} and {\HERWIG}7 using the CH3 tune~\cite{Gieseke:2003rz, Ball:2014uwa, Webber:1983if, Sirunyan:2020pqv}, represented by the pink and orange bands in Figs.~\ref{fig:unfoldedCorr_ModelComp_Cent4}--\ref{fig:unfoldedCorr_ModelComp_Cent1}, respectively, to investigate potential medium-induced jet substructure modifications compared to the perturbative QCD estimation for vacuum fragmentation. The ratios between data and simulation are shown in the lower panels. The comparison presented in Figs.~\ref{fig:unfoldedCorr_ModelComp_Cent4}--\ref{fig:unfoldedCorr_ModelComp_Cent1} shows that the \PYTHIA predictions align well with the measured data for all \pt intervals of the most peripheral events studied, while \HERWIG expects wider distributions at each \pt. Larger discrepancies become evident across all \pt ranges from the centrality interval of 10--30\%, and \HERWIG exhibits a consistently larger deviation from data as compared to \PYTHIA. The level of agreement continues to deteriorate toward more central events, as the data distributions become progressively narrower. 

The prediction from the \JEWEL 2.2.0 model~\cite{jewel}, represented by hatched lines for the 0--10\% centrality interval, is shown in Fig.~\ref{fig:unfoldedCorr_ModelComp_Cent1}. The \JEWEL model is a dynamic framework for jet evolution in a dense background, based on perturbative QCD. However, the model qualitatively fails to describe the 0--10\% centrality collision data across all \pt intervals. While differences exist with and without medium recoil effects, these primarily influence the yield of recoiling jets rather than the distribution shape.
  
The predicted color charge dependence of partonic energy loss should result in a more significant migration of gluon-initiated jets toward lower final-state energy values compared to quark jets. This is because of the higher color charge associated with gluons, which allows them to interact more strongly with the medium~\cite{CMS:2024krd, Zhang:2023oid}. Since the \DRaxis distributions of gluon jets are significantly broader than those of quark jets, differences in energy loss could also lead to the narrowing of jet axis decorrelations. To illustrate this effect, the data are compared to a \PYTHIA-based \DRaxis distribution constructed using predictions from phenomenological models~\cite{mediumqg, mediumqg2}, where medium-induced modifications of jet functions result in changes to the fraction of quark- and gluon-initiated jets in PbPb collisions compared to elementary collisions. Such a model, labeled as ``medium q/g'' in Fig.~\ref{fig:unfoldedCorr_ModelComp_Cent1}, describes the most central data points (for which predictions were available) across the majority of the measured \DRaxis range. At higher \DRaxis in the higher \pt bins, the \PYTHIA-based \DRaxis distributions are above the data, even when considering mostly narrow (quark-initiated) jets in simulation. This suggests that while the bulk of the distribution is well described by the model, the large-\DRaxis behavior cannot be reconciled through changes in quark/gluon composition of the sample alone, indicating the need to account for medium-induced substructure changes.

In Figs.~\ref{fig:unfoldedCorr_ModelComp_Cent4}--\ref{fig:unfoldedCorr_ModelComp_Cent1}, we have added a set of curves labeled ``two-component'', based on a simple model. These curves correspond to a fit of {\PYTHIA}8 with the CP5 tune to the unfolded data, aimed at further testing whether the modified quark/gluon fractions from the model discussed above provide an optimal description of the measured data. In this study, all quark flavors are combined into a single quark template (with different flavor contributions varied for uncertainty evaluations), and a two-component \PYTHIA-based quark/gluon template fit is performed on the fully unfolded measured distributions. This simple model attributes the entire modification to changes in the flavor composition of the sample, thereby providing only a lower limit for the broadest (gluon-like) contribution. The gluon-like jet fraction preferred by the fit aligns with \PYTHIA expectations in the 50--80\% centrality range, but a preference for a much lower fraction is observed in more central events. Notably, the estimated limit on the gluon fraction for 0--10\% central collisions from this simplified  model is in excellent agreement with predictions from the ``medium q/g'' model~\cite{mediumqg, mediumqg2} (as shown in Appendix~\ref{sec:tempFit_gFrac}), as also indicated by the ratio of the \DRaxis distributions with the best fit in the lower panel.
  
\section{Summary}
\label{sec:summary}
This paper presents the first measurement of jet axis decorrelation (\DRaxis) between two types of jet axes for inclusive jets from lead-lead (\PbPb) collision data collected with the CMS detector at the LHC, at a nucleon-nucleon center-of-mass energy of 5.02\TeV. These measurements are performed using anti-\kt jets with a radius parameter $R = 0.4$ and with pseudorapidity $\abs{\eta} < 1.6$ in four centrality intervals and four transverse momentum \pt intervals ranging from 120 to 300\GeV. Significant modifications of \DRaxis distributions are observed in central compared to peripheral collisions, with a progressive narrowing of the distributions toward more central events. This narrowing behavior could result from medium-induced modifications of the internal jet structure, although the possibility of a selection bias towards a less-quenched sample should be considered. Such a bias could arise from the predicted color charge dependence of partonic energy loss, with gluon jets expected to lose more energy than quark jets. The measured distributions from peripheral collisions are described by the {\PYTHIA}8 event generator without any jet quenching effects. Progressive narrowing of the \DRaxis distribution is observed toward more central events for all \pt intervals studied. Comparisons with phenomenological and simple simulation-based models indicate that the observed narrowing in the \DRaxis distributions in central PbPb collisions may be largely described by a decrease in the relative gluon jet abundance, except for higher jet \pt intervals. Deviations at high \pt suggest that accounting for jet substructure modifications or a more sophisticated implementation of color charge effects in energy loss is needed to fully describe the data. These new measurements explore jet substructure in previously unexplored kinematic domains and show great promise for providing new insights into the color charge dependence of energy loss.

\begin{acknowledgments}
  We congratulate our colleagues in the CERN accelerator departments for the excellent performance of the LHC and thank the technical and administrative staffs at CERN and at other CMS institutes for their contributions to the success of the CMS effort. In addition, we gratefully acknowledge the computing centers and personnel of the Worldwide LHC Computing Grid and other centers for delivering so effectively the computing infrastructure essential to our analyses. Finally, we acknowledge the enduring support for the construction and operation of the LHC, the CMS detector, and the supporting computing infrastructure provided by the following funding agencies: SC (Armenia), BMBWF and FWF (Austria); FNRS and FWO (Belgium); CNPq, CAPES, FAPERJ, FAPERGS, and FAPESP (Brazil); MES and BNSF (Bulgaria); CERN; CAS, MoST, and NSFC (China); MINCIENCIAS (Colombia); MSES and CSF (Croatia); RIF (Cyprus); SENESCYT (Ecuador); ERC PRG, RVTT3 and MoER TK202 (Estonia); Academy of Finland, MEC, and HIP (Finland); CEA and CNRS/IN2P3 (France); SRNSF (Georgia); BMBF, DFG, and HGF (Germany); GSRI (Greece); NKFIH (Hungary); DAE and DST (India); IPM (Iran); SFI (Ireland); INFN (Italy); MSIP and NRF (Republic of Korea); MES (Latvia); LMTLT (Lithuania); MOE and UM (Malaysia); BUAP, CINVESTAV, CONACYT, LNS, SEP, and UASLP-FAI (Mexico); MOS (Montenegro); MBIE (New Zealand); PAEC (Pakistan); MES and NSC (Poland); FCT (Portugal); MESTD (Serbia); MICIU/AEI and PCTI (Spain); MOSTR (Sri Lanka); Swiss Funding Agencies (Switzerland); MST (Taipei); MHESI and NSTDA (Thailand); TUBITAK and TENMAK (Turkey); NASU (Ukraine); STFC (United Kingdom); DOE and NSF (USA).
  
  \hyphenation{Rachada-pisek} Individuals have received support from the Marie-Curie program and the European Research Council and Horizon 2020 Grant, contract Nos.\ 675440, 724704, 752730, 758316, 765710, 824093, 101115353, 101002207, and COST Action CA16108 (European Union); the Leventis Foundation; the Alfred P.\ Sloan Foundation; the Alexander von Humboldt Foundation; the Science Committee, project no. 22rl-037 (Armenia); the Fonds pour la Formation \`a la Recherche dans l'Industrie et dans l'Agriculture (FRIA-Belgium); the Beijing Municipal Science \& Technology Commission, No. Z191100007219010 and Fundamental Research Funds for the Central Universities (China); the Ministry of Education, Youth and Sports (MEYS) of the Czech Republic; the Shota Rustaveli National Science Foundation, grant FR-22-985 (Georgia); the Deutsche Forschungsgemeinschaft (DFG), among others, under Germany's Excellence Strategy -- EXC 2121 ``Quantum Universe" -- 390833306, and under project number 400140256 - GRK2497; the Hellenic Foundation for Research and Innovation (HFRI), Project Number 2288 (Greece); the Hungarian Academy of Sciences, the New National Excellence Program - \'UNKP, the NKFIH research grants K 131991, K 133046, K 138136, K 143460, K 143477, K 146913, K 146914, K 147048, 2020-2.2.1-ED-2021-00181, TKP2021-NKTA-64, and 2021-4.1.2-NEMZ\_KI-2024-00036 (Hungary); the Council of Science and Industrial Research, India; ICSC -- National Research Center for High Performance Computing, Big Data and Quantum Computing and FAIR -- Future Artificial Intelligence Research, funded by the NextGenerationEU program (Italy); the Latvian Council of Science; the Ministry of Education and Science, project no. 2022/WK/14, and the National Science Center, contracts Opus 2021/41/B/ST2/01369 and 2021/43/B/ST2/01552 (Poland); the Funda\c{c}\~ao para a Ci\^encia e a Tecnologia, grant CEECIND/01334/2018 (Portugal); the National Priorities Research Program by Qatar National Research Fund; MICIU/AEI/10.13039/501100011033, ERDF/EU, "European Union NextGenerationEU/PRTR", and Programa Severo Ochoa del Principado de Asturias (Spain); the Chulalongkorn Academic into Its 2nd Century Project Advancement Project, and the National Science, Research and Innovation Fund via the Program Management Unit for Human Resources \& Institutional Development, Research and Innovation, grant B39G670016 (Thailand); the Kavli Foundation; the Nvidia Corporation; the SuperMicro Corporation; the Welch Foundation, contract C-1845; and the Weston Havens Foundation (USA).
\end{acknowledgments}

\bibliography{auto_generated}

\clearpage
\appendix
\numberwithin{figure}{section}
\section{Expected gluon fraction in inclusive jets in vacuum and medium}
\label{sec:tempFit_gFrac}
\begin{figure}[htbp!]
\centering
\includegraphics[width=0.60\textwidth]{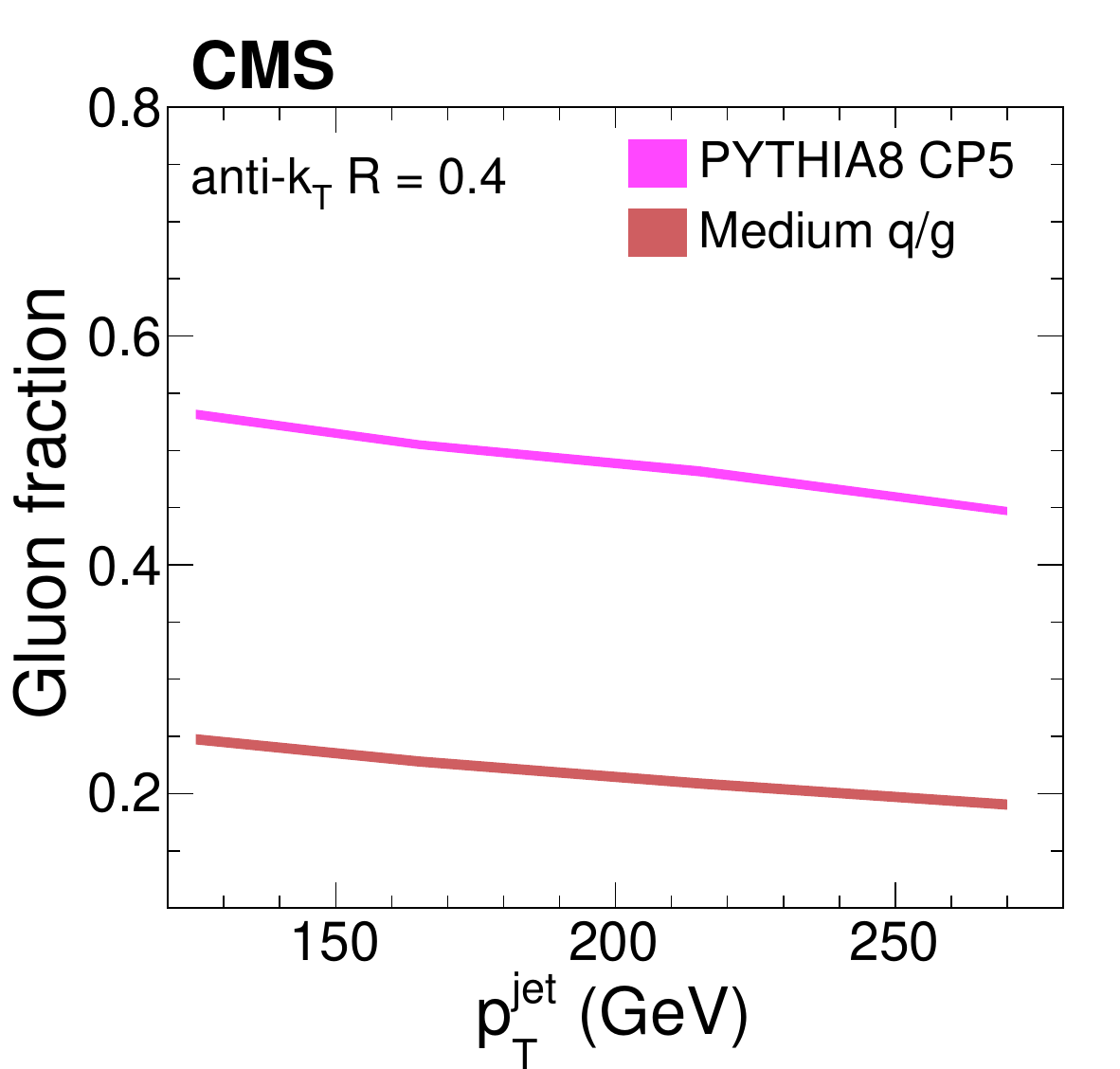}
\caption{Expected gluon fraction in inclusive $R = 0.4$ jet sample in the vacuum (based on {\PYTHIA}8 with tune CP5) and in medium (based on Refs.~\cite{mediumqg, mediumqg2}).}
\label{fig:gluonFrac}
\end{figure}

\cleardoublepage \section{The CMS Collaboration \label{app:collab}}\begin{sloppypar}\hyphenpenalty=5000\widowpenalty=500\clubpenalty=5000
\cmsinstitute{Yerevan Physics Institute, Yerevan, Armenia}
{\tolerance=6000
V.~Chekhovsky, A.~Hayrapetyan, V.~Makarenko\cmsorcid{0000-0002-8406-8605}, A.~Tumasyan\cmsAuthorMark{1}\cmsorcid{0009-0000-0684-6742}
\par}
\cmsinstitute{Institut f\"{u}r Hochenergiephysik, Vienna, Austria}
{\tolerance=6000
W.~Adam\cmsorcid{0000-0001-9099-4341}, J.W.~Andrejkovic, L.~Benato\cmsorcid{0000-0001-5135-7489}, T.~Bergauer\cmsorcid{0000-0002-5786-0293}, S.~Chatterjee\cmsorcid{0000-0003-2660-0349}, K.~Damanakis\cmsorcid{0000-0001-5389-2872}, M.~Dragicevic\cmsorcid{0000-0003-1967-6783}, P.S.~Hussain\cmsorcid{0000-0002-4825-5278}, M.~Jeitler\cmsAuthorMark{2}\cmsorcid{0000-0002-5141-9560}, N.~Krammer\cmsorcid{0000-0002-0548-0985}, A.~Li\cmsorcid{0000-0002-4547-116X}, D.~Liko\cmsorcid{0000-0002-3380-473X}, I.~Mikulec\cmsorcid{0000-0003-0385-2746}, J.~Schieck\cmsAuthorMark{2}\cmsorcid{0000-0002-1058-8093}, R.~Sch\"{o}fbeck\cmsAuthorMark{2}\cmsorcid{0000-0002-2332-8784}, D.~Schwarz\cmsorcid{0000-0002-3821-7331}, M.~Sonawane\cmsorcid{0000-0003-0510-7010}, W.~Waltenberger\cmsorcid{0000-0002-6215-7228}, C.-E.~Wulz\cmsAuthorMark{2}\cmsorcid{0000-0001-9226-5812}
% ...
\par}
\vskip\cmsinstskip
\dag:~Deceased\\
$^{1}$Also at Yerevan State University, Yerevan, Armenia\\
$^{2}$Also at TU Wien, Vienna, Austria\\
% ...
\end{sloppypar}
%%% END EDITABLE REGION %%%
% skeleton_end
\end{document}